\documentclass[pdftex,twocolumn,epjc3]{svjour3}          % twocolumn

\RequirePackage[T1]{fontenc}

\smartqed  % flush right qed marks, e.g. at end of proof

\RequirePackage{graphicx}
\usepackage{latexsym}
\usepackage{amssymb, amsmath, } 
\usepackage{subfig}
\RequirePackage{mathptmx}      % use Times fonts if available on your TeX system
\RequirePackage{flushend}
\RequirePackage[numbers,sort&compress]{natbib}
\RequirePackage[colorlinks,citecolor=blue,urlcolor=blue,linkcolor=blue]{hyperref}

\journalname{Eur. Phys. J. C}

\begin{document}

\title{Hydrostatic equilibrium configurations of neutron stars in a non-minimal geometry-matter coupling theory of gravity}


\author{G. A. Carvalho\thanksref{e1,addr1,addr2}
        \and
        P. H. R. S. Moraes\thanksref{addr2,addr3}
        \and
        S. I. dos Santos Jr.\thanksref{addr2}
        \and
        B. S. Gon\c{c}alves\thanksref{addr2}
        \and
        M. Malheiro\thanksref{addr2}
}

%\thankstext[$\star$]{t1}{Thanks to the title}
\thankstext{e1}{e-mail: araujogc@ita.br}
%\thankstext{e2}{e-mail: magic2@xxx.xx}

\institute{Instituto de Pesquisa e Desenvolvimento (IP\&D), Universidade do Vale do Para\'iba, 12244-000, S\~ao Jos\'e dos Campos, SP, Brazil\label{addr1} \and
  Departamento de F\'isica, Instituto Tecnol\'ogico de Aeron\'autica, S\~ao Jos\'e dos Campos, SP, 12228-900, Brazil\label{addr2}
          \and
          Universidade de S\~ao Paulo, Instituto de Astronomia, Geof\'isica e Ci\^encias Atmosf\'ericas,
 R. do Mat\~ao 1226, Cidade Universit\'aria 
05508-090 S\~ao Paulo, SP, Brazil\label{addr3}
}

\date{Received: date / Accepted: date}

\maketitle

\begin{abstract}
In this work we analyze hydrostatic equilibrium configurations of neutron stars in a non-minimal geometry-matter coupling (GMC) theory of gravity. We begin with the derivation of the hydrostatic equilibrium equations for the $f(R,L) $ gravity theory, where $R$ and $L$ are the Ricci scalar and Lagrangian of matter, respectively. We assume $f(R,L)=R/2+[1+\sigma R]L$, with $\sigma$ constant. To describe matter inside neutron stars we assume a relativistic polytropic equation of state $p=K \rho^{\gamma}$, with $\rho$ being the energy density, $K$ and $\gamma = 5/3 $ being constants. We also consider the more realistic equation of state (EoS) known as SLy4, which is a Skyrme type one based on effective nuclear interaction. We show that in this theory it is possible to reach the mass of massive pulsars, such as PSR J2215+5135, for both equations of state. Also, results for mass-radius relation in GMC gravity are strongly dependent on the stiffness of the EoS.
\end{abstract}

%%%%%%%%%%%%%%%%%%%%%%%%%%%%%%%%%%%%%%%%%%%%%%%%%%%%%%%%%%%
\section{Introduction}\label{sec:int}
%%%%%%%%%%%%%%%%%%%%%%%%%%%%%%%%%%%%%%%%%%%%%%%%%%%%%%%%%%%
It is possible to merger geometry and matter in the same action. An enlightening discussion regarding this question was presented in \cite{ludwig/2015}, in which an action like

\begin{equation}\label{i1}
S=-\frac{\kappa}{2}\int d^{4}x\sqrt{-g}\frac{L^{2}}{R},
\end{equation}
was proposed, with $\kappa=8\pi G/c^{4}$, $G$ the gravitational constant, $c$ the speed of light, $g$ the metric determinant, $L$ the matter lagrangian \footnote{The authors in \cite{ludwig/2015} and others have used the notation ``$\mathcal{L}_m$'' for the matter lagrangian. Here, for the sake of simplicity, we will write the matter lagrangian simply as ``$L$''.} and $R$ the Ricci scalar. 

Interestingly, the dynamics in this theory can only exist in the presence of matter, which, indeed, suggests a deeper link between space-time and matter. The inexistence of dynamics in the absence of matter in a theory of gravity fulfills Einstein's initial proposal of having a gravity theory satisfying Mach's principle \cite{dinverno/1992}. 

It was also shown that the theory in \cite{ludwig/2015} reduces to a special case of the scalar-tensor pressuron theory \cite{minazzoli/2013,minazzoli/2014}.

The theory described from action \eqref{i1} can be seen as a particular case of the well-known $f(R,L)$ theory \cite{harko/2010}, proposed by T. Harko and F.S.N. Lobo. In \cite{harko/2010}, the authors generalized the $f(R)$-type gravity models \cite{nojiri/2011}-\cite{appleby/2007} by assuming that the gravitational lagrangian is given by an arbitrary function of both $R$ and $L$.

The viability of $f(R,L)$ candidates for dark energy was analysed from a dynamical system approach in  \cite{azevedo/2016}. Some constraints were put to $f(R,L)$ theories using the COBE-FIRAS measurement of the spectral radiance of the cosmic microwave background \cite{avelino/2018}. Constraints on $f(R,L)$ gravity were also put via energy conditions in  \cite{wu/2014,wang/2012}. Wormhole solutions were also investigated in the $f(R,L)$ gravity context as one can check \cite{garcia/2010,garcia/2011}. 

Some $f(R,L)$ models do not conserve the energy-mo\-men\-tum tensor and the mechanism responsible for that was said to be a gravitational induced particle production, as it was carefully discussed in  \cite{harko/2015,harko/2014}. 

It is also worth to quote that in Reference \cite{avelino/2018b}, it was indicated that the $f(R,L)$ theories of gravity may be regarded as a subclass of the also well known $f(R,T)$ gravity theories \cite{harko/2011}, for which $T$ is the trace of the energy-momentum tensor. Compact stars have been intensively studied in modified theories of gravity, in particular, the $f(R,T)$ gravity has attracted a lot of researcher's attention \cite{Das/2016,Santos/2019,Maurya/2019,Deb/2018,Deb/2019,Deb/2019b,Zubair/2015,Sharif/2019}. The $f(R,T)$ gravity can be considered a generalization of the well known $f(R)$ theory of gravity. $f(R)$ gravity is one of the most famous modified theory of gravity being much explored in past literature \cite{Rahaman/2012,Zubair/2016,Abbas/2015,Astashenok/2013,Capozziello/2016,Astashenok/2015}. Among the motivations of using modified theories we have: the possibility of surpassing the standard maximum mass limits of compact stars and, thus, get in touch with recent observations of massive pulsars and the indirect evidence of super-Chandrasekhar white dwarfs, and, higher order curvature
corrections or curvature-matter coupling may occur in the extreme gravity regimes inside
a neutron star. However, $f(R)$ and $f(R,T)$ gravity lacks the possibility of getting a very small neutron star radius, and thus, the Buchdal and Schwarzschild radius limits are not surpassed. A non-minimal GMC model is a particular choice with\-in the $f(R,L)$ gravity. The $f(R,L)$ gravity has two major advances in comparison with the other quoted modified theories of gravity: 1) The energy-momentum tensor of the $f(R,L)$ gravity is covariantly conserved by using natural choices for the matter lagrangian and for any choice of the $f(R,L)$ functional, this is frequently a problem within the $f(R,T)$ theory of gravity, 2) the junction with the exterior Schwarzschild solution is easily satisfied with\-in the GMC model in $f(R,L)$ gravity, which in turn is often an issue in $f(R)$ models.

Moreover, in Reference \cite{harko/2013}, the $f(R,L)$ gravity action was generalized by inserting on it a scalar field and a kinetic term, constructed from the gradients of the scalar field. A further model with geometry-matter coupling (from now on referred to as GMC) was proposed by Harko in \cite{harko/2008}. For a review on generalized GMC theories, one can check \cite{harko/2014b}.

GMC models have shown to be able to provide great outcomes when applied to fundamental issues of standard gravity, such as dark matter and dark energy, as one can check Refs.\cite{harko/2010b}-\cite{ms/2017}. The first GMC model was proposed by Nojiri \& Odintsov in \cite{nojiri/2004}, where they apply a GMC theory to explain the cosmic acceleration. A GMC model was also considered for solar-like stars in \cite{bertolami/2007,bertolami/2008}, where authors used the Newtonian limit to obtain numerical results for hydrostatic equilibrium of this type of stars and also to derive constraints to the GMC model. 

Here in this work, instead, we will be concerned to the outcomes of applying a GMC model for obtaining the hydrostatic equilibrium configurations of neutron stars (NSs). NSs are supernova remnants known for their high density, strong gravitational field and rapid rotation rate. NS binary systems are among the leading gravitational wave sources \cite{phinney/1991,agathos/2015} and have, indeed, been already detected by Advanced LIGO and Advanced Virgo detectors \cite{abbott/2017}.

Although most NSs have masses $\sim1.3-1.4$M$_\odot$ \cite{lattimer/2012}, there is ample observational support for NSs with  greater masses ($\sim2M_\odot$) \cite{demorest/2010,antoniadis/2013}, which causes some controversy regarding the NSs origin, equation of state (EoS) or even the underlying gravitational theory.

It is worth to remark that in a recent article \cite{Maurice/2019}, it was concluded that the GW 170817 event, coming from a binary system with $2.74^{+0.04}_{-0.01}M_\odot$, may have  resulted in a super-massive magnetar. Also, in \cite{linares/2018} a pulsar with $2.27^{+0.17}_{-0.15}M_\odot$ was reported, being the most massive NS already detected, named PSR J2215+5135. If NSs in $f(R,L)$ gravity can attain this value for the mass it will be the first time that such an object is theoretically predicted in alternative gravity. It also worth to cite that hydrostatic equilibrium configurations of compact were never performed before for the GMC theory considered in this work. 

%It should be highlighted that the existence of PSR J2215+5135 has ruled out a number of E'soS describing nuclear matter inside NSs from General Relativity perspective. On this regard, one could check References \cite{m.dutra/2016}-\cite{m.dutra/2012}, in which robust constraints were put to models of relativistic and non-relativistic nuclear matter. Such an  analysis allowed the inclusion of twelve models describing NSs in the range $[1.93-2.05]M_\odot$ for their maximum masses and two others surpassing this limit, but by necessarily including  hyperons.

%The PSR J2215+5135 is a millisecond pulsar with a rotation period of 2.61ms, located at $\sim 4$kpc from us. It belongs to a specific class of pulsars named \emph{redbacks}, which are binary systems where the companion star has masses in the interval $\sim0.1-0.4$M$_\odot$ \cite{Broderick/2016}. For PSR J2215+5135 the companion star has $0.33^{+0.03}_{-0.02}$M$_\odot$, being classified as a spectral type G5 of the main sequence.
		
In this work we will investigate the possibility of predicting the above pulsar by altering the underlying gravitational theory. Particularly, we will investigate the hydrostatic equilibrium configurations of NSs, with particular equations of state, from a non-minimal GMC model which shall be presented in Section \ref{sec:gmc}. In Section \ref{sec:hee} we will derive the hydrostatic equilibrium equations for the concerned theory. In Section \ref{sec:eos} we will present the EoS that we shall consider for numerically solving the hydrostatic equilibrium equations in Section \ref{sec:ns}. We will discuss our results in Section \ref{sec:dis}.

\section{A non-minimal geometry-matter coupling}\label{sec:gmc}

Here we will work with a GMC theory named $f(R,L)$ gravity \cite{harko/2010}, whose action reads

\begin{equation}\label{gmc1}
S=\int d^{4}x\sqrt{-g}f(R,L),
\end{equation}
with $f(R,L)$ being a function of $R$ and $L$ and $8\pi G$ and $c$ are taken as $1$ from now on. One can note from \eqref{gmc1} that when $f(R,L)=R/2+L$, the usual Einstein-Hilbert action is retrieved, such that the variational principle application implies the usual Einstein's field equations $G_{\mu\nu}=T_{\mu\nu}$, with $G_{\mu\nu}$ being the Einstein tensor and $T_{\mu\nu}$ the energy-momentum tensor.

By following references \cite{garcia/2010,garcia/2011}, we will consider the case $f(R,L)=f_1(R)/2+[1+\sigma f_2(R)]L$, with $f_1(R)$ and $f_2(R)$ being functions of $R$ only and the parameter $\sigma$ can be said to characterize the strength of the coupling. 
For the sake of simplicity we shall take $f_1(R)=f_2(R)=R$. Also, we assume $L=-p$ \cite{harko/2014b}, with $p$ being the pressure of the fluid considered. Taking all these considerations into account, the variational principle applied to \eqref{gmc1} yields the following field equations \cite{harko/2010}

\begin{eqnarray}\label{gmc3}
&&(1-2\sigma p)G_{\mu\nu}+\frac{1}{3}Rg_{\mu\nu}-\frac{\sigma p}{3}Rg_{\mu\nu}=\nonumber\\
&&(1+\sigma R)\left(T_{\mu\nu}-\frac{1}{3}Tg_{\mu\nu}\right)-2\sigma \nabla_\mu\nabla_\nu p.
\end{eqnarray}

Moreover, the covariant derivative of the ener\-gy-mo\-men\-tum tensor reads \cite{harko/2010}

\begin{equation}\label{gmc4}
\nabla^{\mu}T_{\mu\nu}=(-pg_{\mu\nu}-T_{\mu\nu})\nabla^{\mu}\ln(\sigma R). 
\end{equation}

One may wonder, in another perspective, about the cosmological consequences of Eq.(3). In fact, a cosmological model derived from the substitution of the Fried\-mann-Lem\^ai\-tre-Robertson-Walker metric as well as the energy-momen\-tum tensor of a perfect fluid in Eq.(3)  has not been reported in the literature so far, but shall be soon. For now, we can mention that other GMC models have already been applied to cosmology, yielding well behaved scenarios \cite{ms/2017,zaregonbadi/2016,nesseris/2009,wang/2013}.

\section{The hydrostatic equilibrium equations in a non-minimal geometry-matter coupling model}\label{sec:hee}

The hydrostatic equilibrium equations in the concerned GMC theory will be obtained from the substitution of the static spherically symmetric metric 

\begin{equation}\label{hee1}
ds^{2}=e^{\alpha(r)}dt^2-e^{\beta(r)}dr^{2}-r^{2}(d\theta^{2}+\sin^{2}\theta d\phi^{2}),
\end{equation}
in \eqref{gmc3}-\eqref{gmc4}, with $\alpha(r)$ and $\beta(r)$ being metric potentials depending on $r$ only.

From the substitution of \eqref{hee1} into \eqref{gmc4} and assuming the energy-momentum tensor of a perfect fluid, such that $T_{\mu\nu}=\texttt{diag}(e^\alpha\rho,e^\beta p,r^2p,r^2\sin^2\theta p)$, with $\rho$ being the matter-ener\-gy density, we have for $\nu=0$%we can conclude that the energy-momentum tensor is always covariantly conserved independently of the functional form assumed for $f(R,L)$.

\begin{eqnarray}
  \nabla^{\mu}T_{\mu 0}&=& (-pg_{\mu 0}- T_{\mu 0}) \nabla^\mu {\rm ln} (\sigma R),\nonumber \\
  \nabla^{\mu}T_{\mu 0}&=& (-pg_{0 0}- T_{0 0}) \nabla^0 {\rm ln} (\sigma R),
\end{eqnarray}
and considering that we are concerned to a static case, we have $\nabla^0 {\rm ln} (\sigma R)=0$. Now, for $\nu=i$, where $i=1$, 2 and 3, we have
\begin{equation}\label{nab:1}
  \nabla^{\mu}T_{\mu i}= (-pg_{\mu i}- T_{\mu i}) \nabla^\mu {\rm ln} (\sigma R),
\end{equation}
and the right side of \eqref{nab:1} provides the term $(-pg_{i i}- T_{i i})$, which is identically null for all values of $i$. This implies that the energy-momentum tensor is covariantly con\-ser\-ved, i.e.,
\begin{equation}\label{conserv}
  \nabla^{\mu}T_{\mu\nu}=0,
\end{equation}
which justifies our choice of matter Lagrangian as $L=-p$, since a different choice of Lagrangian would give a non-con\-served energy-momentum tensor.

The $00$ and $11$ components of the field equations \eqref{gmc3} for metric \eqref{hee1} read, respectively,

\begin{eqnarray}\label{eq:camp:1}
  &&\frac{(1-2\sigma p)}{r^2}\left(r-re^{-\beta}\right)' +(1-\sigma p)\frac{R}{3}\nonumber\\
  &&=(1+\sigma R)\left(\frac{2}{3}\rho+p\right)+\sigma e^{-\beta}\alpha' p',
\end{eqnarray}
\begin{eqnarray}\label{eq:camp:2}
&&\frac{(1-2\sigma p)}{r^2}\left(e^{-\beta}-1+e^{-\beta}\alpha' r\right) +(\sigma p-1)\frac{R}{3}\nonumber\\
 &&=(1+\sigma R)\frac{\rho}{3}-2\sigma e^{-\beta}\left(p''-\frac{\beta'}{2}p'\right),
\end{eqnarray}
with primes denoting derivatives with respect to the radial coordinate $r$.

The conservation of the energy-momentum tensor \eqref{conserv} yields to

\begin{equation}\label{tov}
  p'=-(\rho+p)\frac{\alpha'}{2}.
\end{equation}

As the Ricci scalar becomes a degree of freedom, another equation can be derived from the trace of the field equations as

\begin{equation}\label{trace}
(1+2\sigma p)R=-(1+\sigma R)T-6\sigma \Box p,
\end{equation}
where the D'Alambertian operator reads
\begin{equation}
  \Box=-e^{-\beta}\left[\frac{d^2}{dr^2}-\frac{\beta'}{2}\frac{d}{dr}+\frac{\alpha'}{2}\frac{d}{dr}+\frac{2}{r}\frac{d}{dr}\right].
\end{equation}

In order to obtain the hydrostatic equilibrium configurations, Equation \eqref{trace} needs to be included in the system of differential equations \eqref{eq:camp:1}, \eqref{eq:camp:2} and \eqref{tov}. The unknowns are $R$, $\alpha$, $\beta$, $\rho$ and $p$ so that we have five variables and four equations. In this way, we need to define a relation between $p$ and $\rho$, namely an EoS.

\section{The equation of state for nuclear matter inside neutron stars}\label{sec:eos}

\begin{figure}[h!]
\centering
\subfloat[Pressure vs. energy density.\label{eos}]{\includegraphics[width=0.4\textwidth]{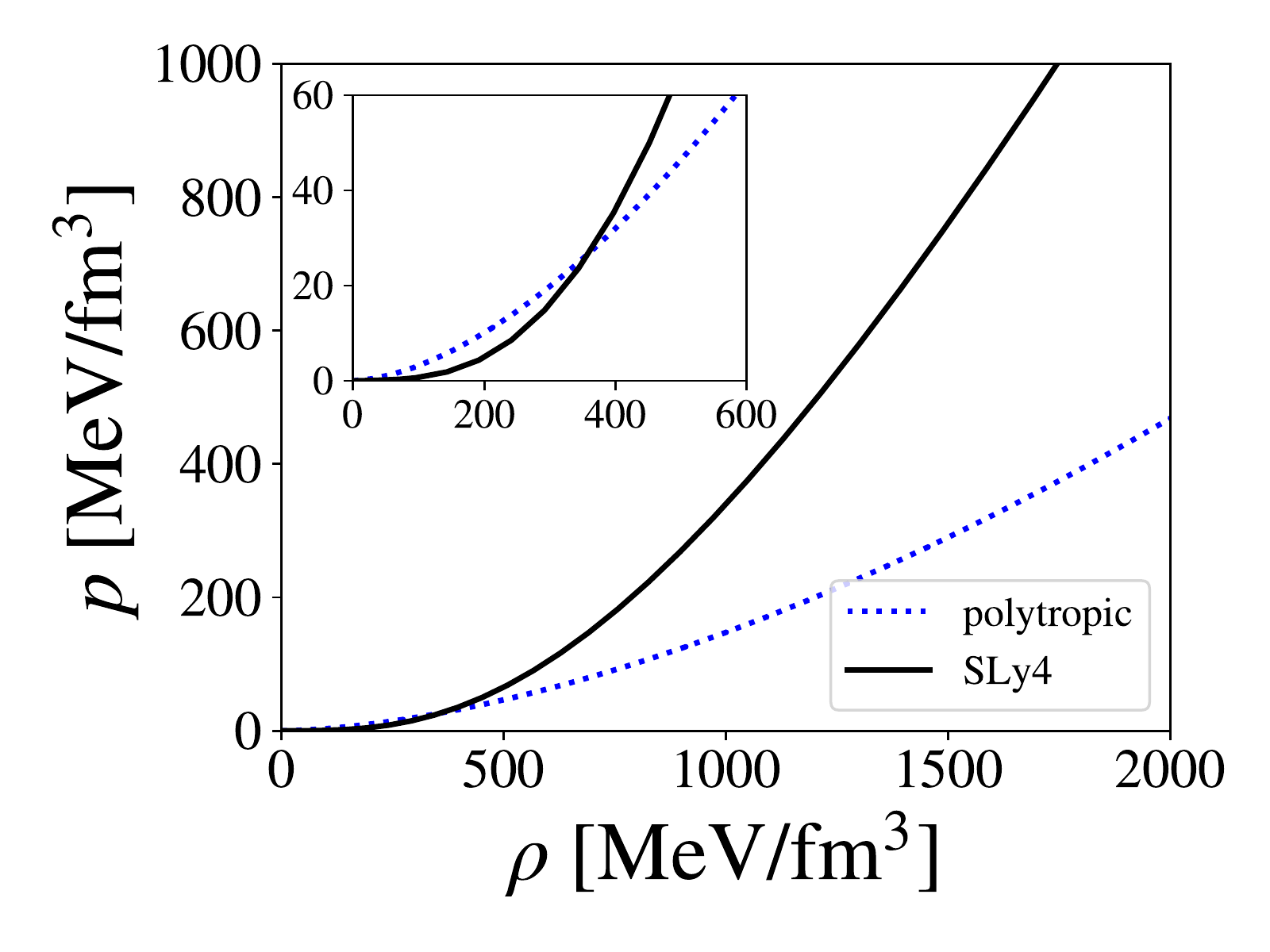}}\hfill
\subfloat[Sound velocity.\label{soundv}]{\includegraphics[width=0.4\textwidth]{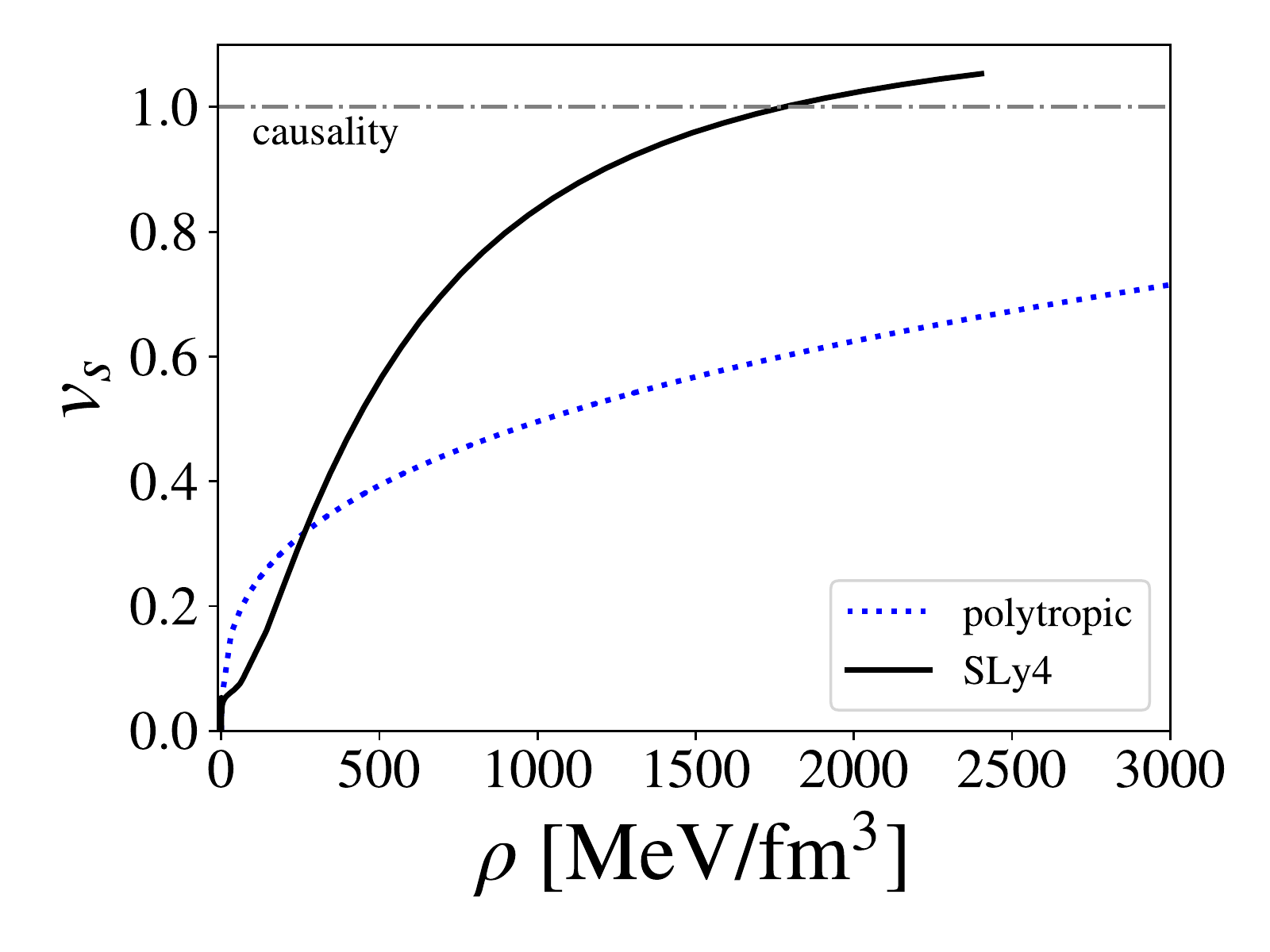}}
\caption{Equation of state properties. Sound velocity, $v_s$, must respect the causal limit, which states $v_s<1$. Bottom panel shows that polytropic EoS does not violate causality, while SLy4 does violate for densities slightly above the star central density for the maximum mass star in GR (see Fig. \ref{massdensitySLy}).} \label{fig:eos}
\end{figure}

Once provided an EoS for the matter inside the star, Eqs.\eqref{eq:camp:1}-\eqref{trace} can be numerically solved. One of the most simple and oftenly used NS EoS in the literature is the polytropic one. Following References \cite{tooper/1964}-\cite{moraes/2016}, the relation between $p$ and $\rho$ can be regarded as $p=k\rho^{\gamma}$, where $k$ is constant and $\gamma=1+1/n$, where $n$ is the so-called polytropic index. In this work we consider $k=1.475\times 10^{-3} ({\rm fm^3/MeV})^{2/3}$ and $n=3/2$. Although this kind of equation of state does not represents the state of the art to describe neutron star microphysics, several recent works has used this type of equation of state to study compact stars and, in particular, neutron stars in modified theories of gravity \cite{Sharif/2019,Nari/2018,Sharif/2019b,Yousaf/2018}. The main reason for this is that the particular interest is to investigate the modified theory imprints itself. Polytropic approximations are also often used to describe neutron star EoS by fitting of its parameters. It is worth to cite that, here we are dealing with a polytropic relation between pressure and energy density \cite{herrera/2004,herrera/2013}. This kind of relation is way more efficient to fit the equation of state of neutron stars than a polytropic relation between pressure and rest mass density, and, thus, it is also more efficient to describe the neutron star macroscopic features, such as mass-radius relation. The so called SLy equation of state, described in \cite{Douchin/2001}, is a Skyrme type one based on nuclear interactions, and it will also be considered in this work. Both EoS are depicted in Fig.\ref{fig:eos}, where panel \ref{eos} shows pressure as a function of energy density and panel \ref{soundv} presents sound velocity $ v_s = (v_{sound}/c)$  also as function of energy density. From panel \ref{eos} we can see that the polytropic EoS is softer than SLy one, what is the main reason behind the difference on maximum masses in GR, where the polytropic EoS has a maximum mass of $\sim 1.4M_\odot$ and the SLy $\sim 2.0M_\odot$. However, both maximum masses in GR are not capable to describe the PSR J2215+5135 mass even if one consider the lower bound. In panel \ref{soundv}, the constrain on sound velocity, $v_s<1$, is not respected for the SLy EoS for densities larger than $\sim 1800$ MeV/fm$^3$ (or $\approx 12.2 \rho_0$, where $\rho_0$ represents the saturation nuclear  energy density), whe\-reas polytropic equation of state violates causality only for densities larger than $8200$ MeV/fm$^3$ ($55.8\rho_0$).

\section{Numerical procedure, mass-radius and mass-central density relations}\label{sec:ns}

The system of differential equations previously derived are solved here considering the usual initial and boundary conditions, i.e., the pressure and density have an initial value at the center of the star, namely, $p(0)=p_C$ and $\rho(0)=\rho_C$, and the equations are integrated until the pressure vanishes, that is, $p(r=R_{\star})=0$, where $R_{\star}$ is the radius of the star \cite{carvalho/2018}. It is worth to quote that $R$ is also zero at the surface according to \eqref{trace} and, hence, the junction with the exterior Schwarzschild solution is satisfied. The initial value of $\beta$ is considered to be null, in analogy with the interior Schwarzschild solution. The initial value of $\alpha$ is arbitrary since the differential equations depend only on its derivative $\alpha'$. 

\begin{figure}[h!]
\centering
\subfloat[Mass-radius relation for polytropic stars.\label{massradius}]{\includegraphics[width=0.4\textwidth]{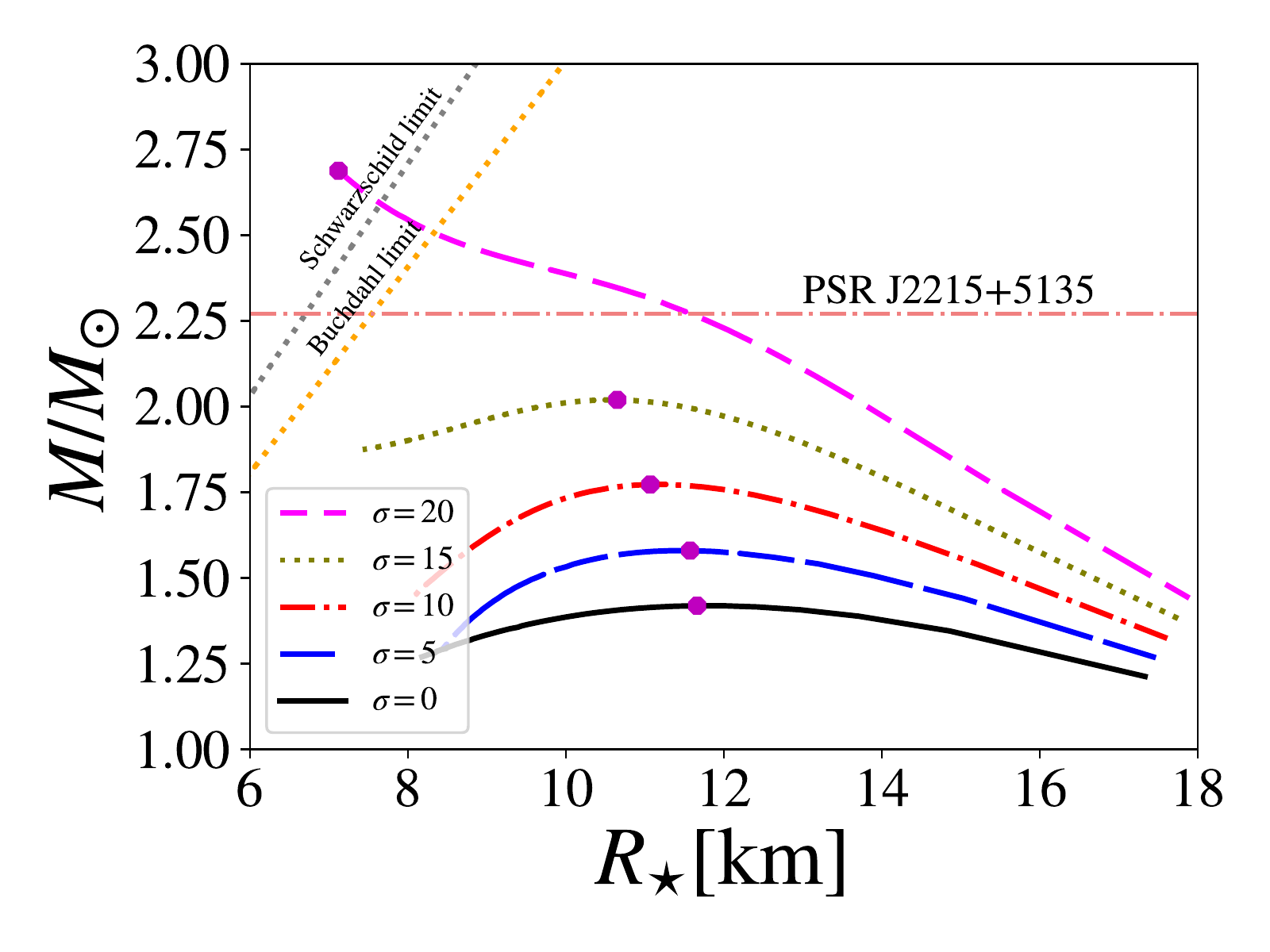}}\hfill
\subfloat[Mass-central density relation for polytropic stars.\label{massdensity}]{\includegraphics[width=0.4\textwidth]{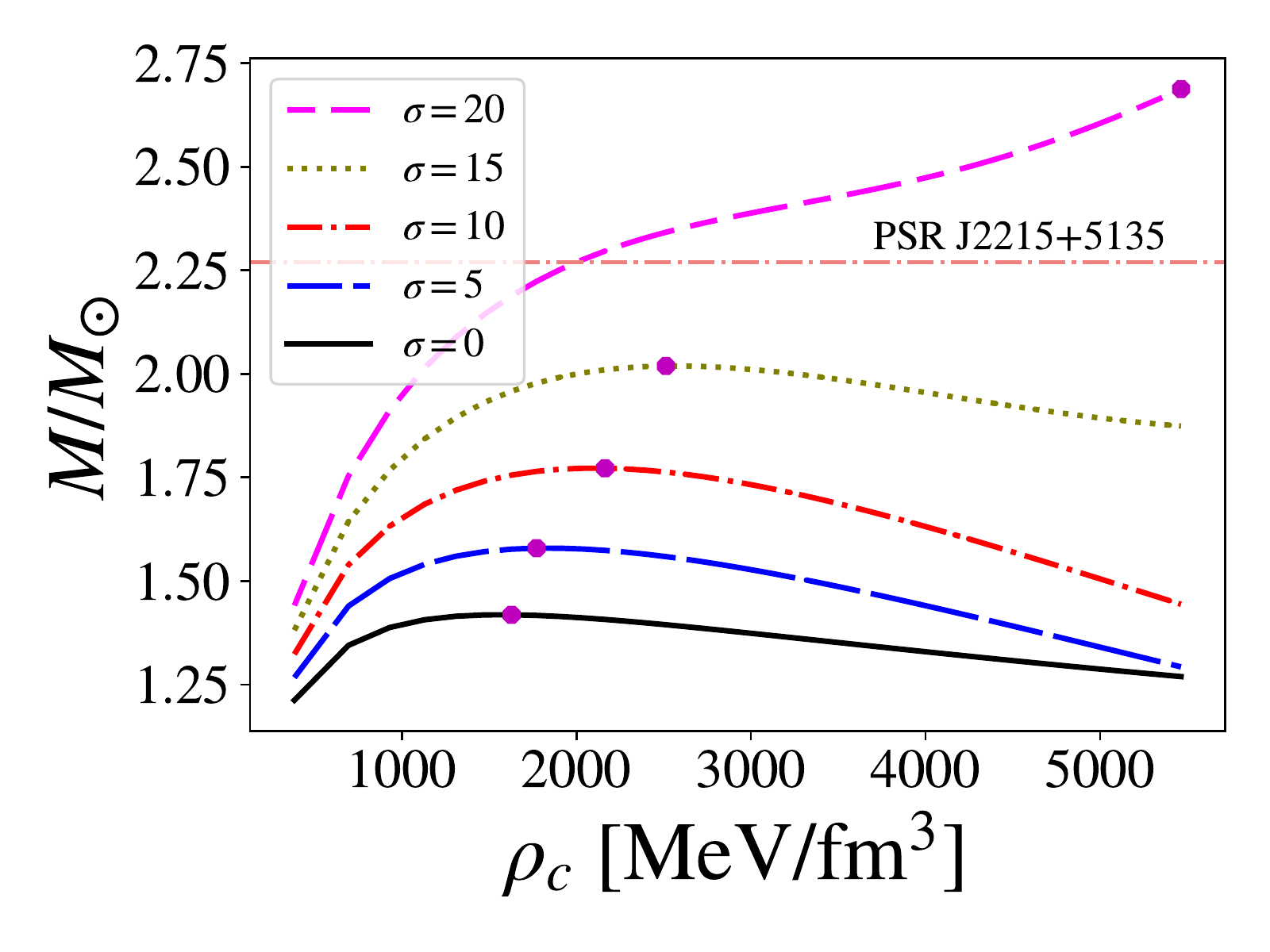}}
\caption{(color online) Mass-radius \eqref{massradius} and mass-central density \eqref{massdensity} relations for several values of $\sigma$ for the polytropic EoS. The magenta circles mark the maximum mass on each curve. In panel \eqref{massradius} the Buchdahl and Schwarzschild limits are indicated by an orange-dotted and a gray-dotted line, respectively.} \label{fig:1}
\end{figure}

It should also be cited that since the energy-mo\-men\-tum tensor is covariantly conserved (see Eq. \eqref{conserv}), the total stellar mass can be calculated in its standard form as

\begin{equation}
M=\int_0^{R_\star} 4\pi r^2 \rho(r)dr.
\end{equation}

In Figure \ref{massradius} we present the mass-radius relation for the polytropic neutron stars in the $f(R,L)=R/2+L+\sigma R L$ theory of gravitation for different values of $\sigma$, the coupling parameter. Magenta circles mark the maximum masses for each value of  $\sigma$ and we also present the lines of the Buchdahl and Schwarzs\-child radius limits, and show that these limits can be surpassed in the GMC theory. From this figure we can see that the radius of the neutron star ranges from 7 to 18km, which is within the expected values of neutron star radii from observational constraints \cite{a.mariano/2018,most/2018}. On the other hand, the maximum mass of the polytropic neutron star is largely affected, being able to reach values up to $\sim$2.69$M_\odot$.

\begin{figure}[h!]
\centering
\subfloat[Mass-radius relation for the SLy equation of state.\label{massradiusSLy}]{\includegraphics[width=0.4\textwidth]{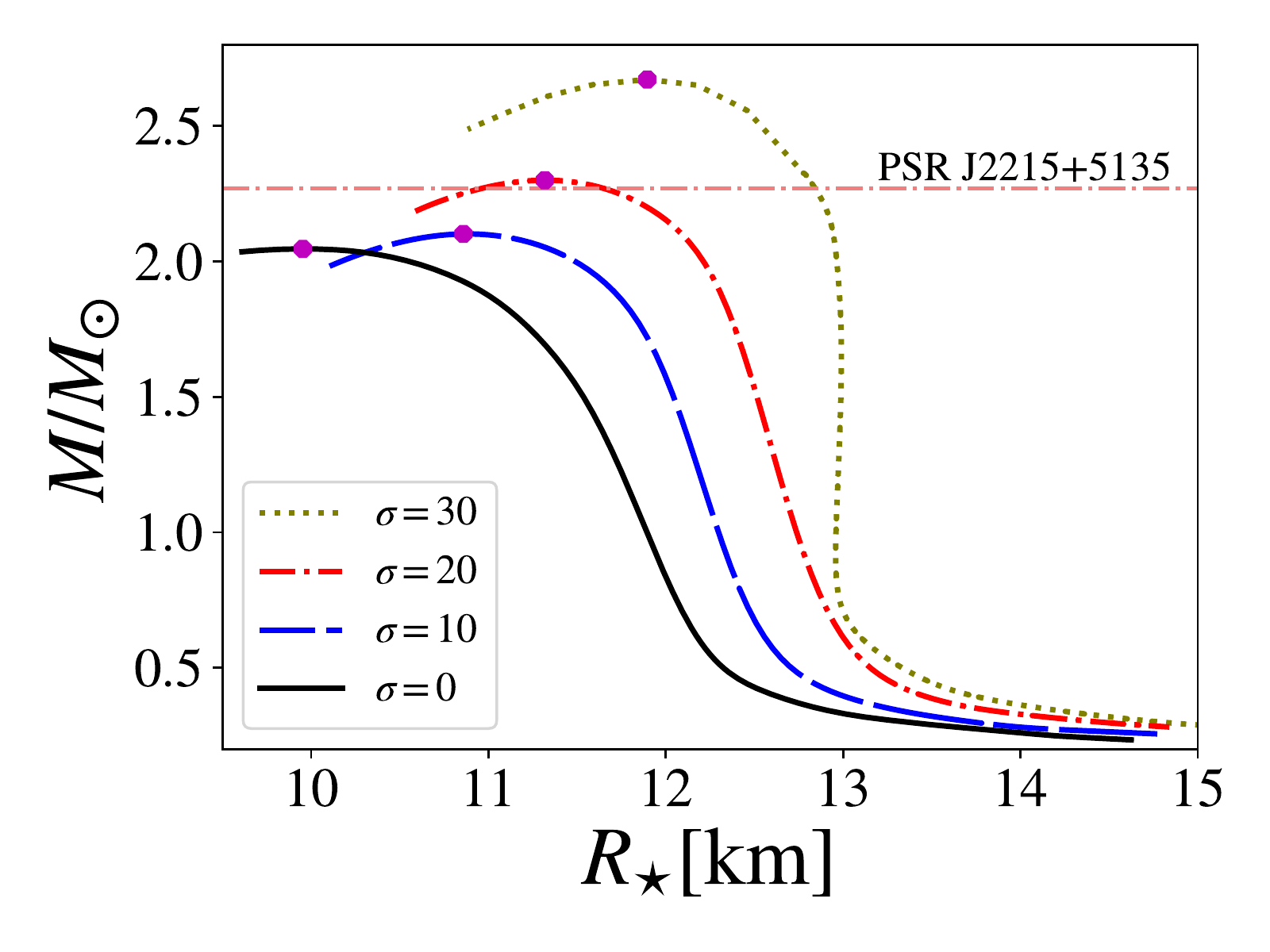}}\hfill
\subfloat[Mass-central density relation for the SLy equation of state.\label{massdensitySLy}]{\includegraphics[width=0.4\textwidth]{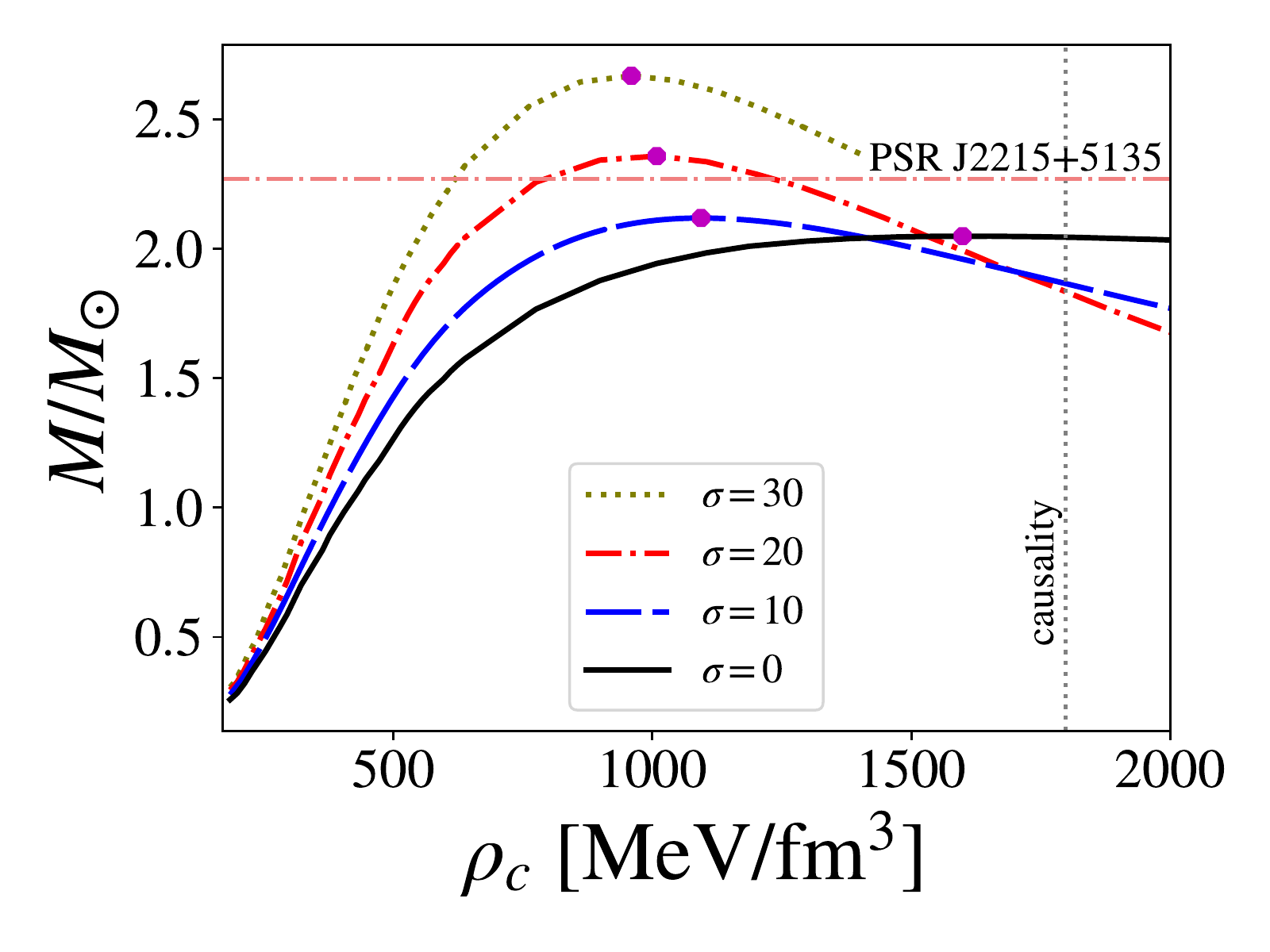}}
\caption{(color online) Mass-radius \eqref{massradiusSLy} and mass-central density \eqref{massdensitySLy} relations for several values of $\sigma$ and for the SLy4 EoS. The magenta circles mark the maximum mass on each curve. In both panels a horizontal dashed-dotted line indicates the PSR J2215+5135 mass. In panel \ref{massdensitySLy} vertical dotted line represents the causality threshold.} \label{fig:2}
\end{figure}

Furthermore, in Fig. \ref{fig:2} we present the mass-radius and mass-central density relation for the Sly EoS \cite{Douchin/2001}. From panel \ref{massradiusSLy}, we observe that stars become more massive according $\sigma$ is increased. In particular, maximum mass increases from $~2M_\odot$ ($\sigma=0$) to $2.6M_\odot$ ($\sigma=30$), which represents a $\sim 30\%$ increasing. In addition, for the SLy equation of state the maximum mass points are attained for larger values of radius. This is due the maximum mass points are being obtained for smaller values of central energy density, see Fig. \ref{massdensitySLy}. This means that the GMC model yields more massive and larger stars for the SLy EoS, and thus, in this case, the Buchdahl and Schwarzschild limits are not surpassed. This opposite behavior from one EoS to another could help to differ results from different EoS in GMC gravity once given observational values of gravitational redshift. The main reason for this opposite scenario is that used equations of state are quite non-similar. For example, as stated before, the Sly EoS is almost non-causal, while polytropic one is far from being causal for the considered density interval. This means that the larger the softness possibly the larger the sensibility of results concerning to changes in parameter $\sigma$. It is worth to cite that further investigations analyzing different EoSs could help to better understand this phenomena.

It is worth noticing that the behavior of the mass-radius relation for polytropic NS changes according to the value of $\sigma$ in a way that a change of concavity is present for $\sigma\geq 15$. Due to this effect, in the curve for $\sigma=20$ the mass does not decrease with the reduction of the stellar radius, i.e., $dM/dR$ is always negative.

A similar behavior is presented in the mass-central density relation of Figure \ref{massdensity} above, where the mass does not decrease with the increasing on central energy density. It is worth to cite that, {\it a priori}, this behavior does not represent an instability on the NSs and hence a maximum stable mass cannot be set only by means of equilibrium configurations. In this sense, the GMC model can predict very high mass NSs, such as the pulsar PSR J2215+5135, for instance, with a very simple and soft EoS. Those stars also present small radius, such as 8km. Such a combination of high mass and small radius can overcome the Buchdahl limit ($GM/c^2R<4/9$) and also the Schwarzschild radius limit for the star to become a black hole, indicating that the Schwarzschild radius and Buchdahl limit are changed in the GMC model (see also Figure \ref{massradius}). We stress that in order to establish a maximum mass limit within the GMC model one needs to perform a study about radial oscillations \cite{friedman/1988}-\cite{domingo/2015}. We plan to address all those stability issues in a forthcoming work. 
However, most massive polytropic neutron stars obtained here have a huge central energy density, as we already pointed out before,  surpassing the saturation nuclear density in a few dozens, which may be unrealistic as we are working with a simple EoS treatment. This indicates that, even polytropic stars respecting causality and trace positiveness, the huge maximum mass may not be physically feasible since they are obtained for very high values of central energy density. In addition, when we took the SLy equation of state a remarkable increasing on the maximum mass is presented. For the SLy EoS, we also noticed that maximum central energy densities are smaller than that obtained in GR, which means causality is always respected and the mass of the PSR J2215+5135 is attainable for $\sigma \sim 20$ with maximum central energy density of $\approx 1000$ MeV/fm$^3$ ($\approx 6.8\rho_0$), which is a more feasible scenario.

Anyhow, a maximum limit for the GMC parameter $\sigma$ can be established from observations concerning the pulsar PSR J2215+5135. For the considered range of central pressure, the value that describes the mass of such a pulsar is about $\sigma=18$ for the polytropic EoS and about $\sigma\approx 20$ for the SLy one. The maximum mass for
$\sigma=20$ is $M_{\rm max}=2.69M_\odot$, which represents a remarkable increasing of $\sim 90$\% with respect to maximum mass point of General Relativity ($\sigma=0$) concerning the polytropic EoS.

\section{Discussion}\label{sec:dis}

It has been shown that it is possible to avoid the Big-Bang singularity through a GMC model \cite{banados/2010}. In such an approach, the authors used the so-called Eddington-inspired Born-Infeld gravity.  The achievement of evading the Big-Bang singularity was expected to be attained only through quantum gravity and, in this sense, GMC gravity models can figure as a great alternative until we derive the ultimate theory of quantum gravity. For more insights on GMC gravity, one can also check \cite{delsate/2012}.

Let us briefly visit some other recent GMC proposals and their applications. The first GMC model was proposed in \cite{nojiri/2004}. In \cite{harko/2018}, the authors presented an extension of teleparallel gravity \cite{cai/2016,bahamonde/2015}, in which the non-metricity is non-minimally coupled to the matter lagrangian. Some cosmological models were derived from such a formalism and have featured an accelerated expansion for late times. In \cite{harko/2010b}, it was shown that the effective energy-mo\-men\-tum tensor of a GMC theory is, indeed, more general than the usual perfect fluid energy-momentum tensor of General Relativity, and that the referred extra terms could be related to  elastic stresses in the body, or to other forms of internal energy. In fact, they could also be related to fluid imperfections, such as viscosity and anisotropy. The field equations for a GMC model by using the Palatini formalism were derived in \cite{harko/2011b}. Furthermore, in \cite{bertolami/2010} it was shown that the dark matter effects can simply be a consequence of GMC and in \cite{bertolami/2011} it was shown that a generalized GMC is compatible with Starobinsky inflation \cite{starobinsky/1982}.

In the present work, we have obtained hydrostatic equilibrium configurations of NSs in a non-minimal GMC gravity model. The underlying gravitational theory was chosen to be the $f(R,L)$ theory \cite{harko/2010}. It is the first time in the literature that the hydrostatic equilibrium equations are solved for neutron stars in a fully relativistic treatment in such a theory. In \cite{bertolami/2007,bertolami/2008} a detailed study was performed for solar-like stars using a similar equation of state as the one polytropic one employed here. However, their equation of state is written in terms of the baryonic mass density instead of energy density, which yields a very different relation between pressure and energy density as showed by \cite{herrera/2004,herrera/2013}. We also consider in the present work the case of SLy EoS , a more realistic and stiff equation of state (EoS) derived from a Skyrme type force based on effective nuclear interaction. 

The main motivation for doing so is related to some recent observations of massive pulsars \cite{demorest/2010,antoniadis/2013}. In fact, even the detection of some super-Chandrasekhar white dwarfs \cite{howell/2006,scalzo/2010} provides a field of investigation for alternative gravity \cite{clmaomm/2017,das/2015}. More remarkably, in \cite{linares/2018}, a pulsar with $\sim2.27M_\odot$, na\-med PSR J2215+5135, was reported as the most massive NS already observed. The achievement of such a mass scale with a very simple EoS was the main motivation of our work, that is, can such a mass scale be predicted through a GMC gravity model with a very simple, and often disregarded in GR, EoS? The answer for the this question is yes, as we explain in the following.  

Fig.\ref{massradius} shows the mass-radius relation of polytropic NSs in the GMC model for different $\sigma$, which features the strength of the GMC. It is clear that for stronger couplings ($\sigma\geq18$), the PSR J2215+5135 mass scale is attained. The mass scale of other massive pulsars, such as J1614-2230 \cite{demorest/2010} and J0348+0432 \cite{antoniadis/2013}, which is $\sim2M_\odot$, is attained for  slightly weaker couplings ($\sigma\sim15$). In addition, the Sly EoS can also achieve the mass of PSR J2215+5135 for $\sigma\sim 20$,  in spite of its maximum in GR ($M_{\rm max}=2.05M_\odot$).

Moreover, the equilibrium configurations of neutron stars in the GMC theory was showed to largely depend on the EoS stiffness. This aspect of GMC may be clarified if further investigations on the GMC theory with other EoS are performed. For now, we can conclude only that for softer EOS the effect of the coupling between background gravity and matter in GMC generates an effective geometry-matter pressure that is smaller when this coupling is absent, and as a consequence a higher central energy density in the matter pressure is needed to compensate gravity and stabilize the star turning it more compact. For stiff EOS, we have the usual feature of $f(R,T)$ or $f(R)$ theories, where the geometry-matter coupling produces an effective  pressure that is stronger and a smaller central energy density in the matter pressure is enough to stabilize the star making it bigger and less dense.

%\acknowledgments
\begin{acknowledgements}
GAC thanks Coordena\c{c}\~ao de Aperfei\c{c}oamento de Pessoal de N\'ivel Superior processes PDSE/88881.188302/2018-01 and PNPD/88887.368365/2019-00. PHRSM thanks S\~ao Paulo Research Foundation (FAPESP), grants 2018/20689-7 and 2015/08476-0, for financial support. The authors also thank FAPESP under the thematic project 13/26258-4. Authors are also in debit with R.V. Lobato for discussions and data.
\end{acknowledgements}

\end{document}